\documentclass{article}
\newcommand{\be}{\begin{equation}}
\newcommand{\en}{\end{equation}}
\newcommand{\bea}{\begin{eqnarray}}
\newcommand{\ena}{\end{eqnarray}}

\def\r{\rho}

\def\eq{equation}
\def\en{eqnarray}
\begin{document}
\begin{titlepage}
\centering{\bf\huge{Cosmic acceleration from interaction of
ordinary fluids}} \\
\vspace{0.5cm} \centering{Nelson Pinto-Neto \footnote{e-mail
address: nelsonpn@cbpf.br}
\\{\small\it{ICRA - Centro Brasileiro de Pesquisas F\'{\i}sicas --
CBPF, \\ Rua Xavier Sigaud, 150, Urca, CEP 22290-180, Rio de
Janeiro, Brazil}} \\

~

Bernardo M.O. Fraga \footnote{e-mail address: bernardo@cbpf.br} \\
{\small\it{Observat\'orio do Valongo - Universidade Federal do Rio
de Janeiro\\ Ladeira Pedro Ant\^onio, 43, Sa\'ude, CEP 20080-090,
Rio de Janeiro, Brazil}}} \vspace{2.0cm}
\par We would like to thank CNPq of Brazil for financial support. One of us
(NPN) would like to thank the French/Brazillian cooperation
CAPES/COFECUB for partial financial support. We would also like to
thank `Pequeno Seminario' of CBPF's Cosmology Group for useful
discussions.

\end{titlepage}

\begin{abstract}
Cosmological models
with two interacting fluids, each satisfying the strong energy condition, are studied
in the framework of classical General Relativity. If the interactions are phenomenologically
described by a power law in the scale factor, the two initial interacting
fluids can be equivalently substituted by two non interacting effective fluids, where one of them may violate
the strong energy condition and/or have negative energy density. Analytical solutions
of the Friedmann equations of this general setting are obtained and studied. One may have,
depending on the scale where the interaction becomes important,
non singular universes with early accelerated phase, or singular models with transition from decelerated
to accelerated expansion at large scales. Among the first, there are bouncing models where contraction
is stopped by the interaction. In the second case, one obtains dark energy expansion rates without dark energy,
like $\Lambda$CDM or phantomic accelerated expansions without cosmological constant or phantoms, respectively.

\vspace{0.3cm}
PACS number(s): 98.80.Cq, 98.80.-k, 98.80.Bp

\vspace{0.3cm} Key words: bouncing universe; interacting fluids; cosmic acceleration

\end{abstract}

\pagebreak
\section{Introduction}

In general, the fluids describing the matter content of
a cosmological model are considered to be non interacting.
For instance, in the Standard Cosmological Model, cold dark matter, neutrinos,
dark energy and the baryon-photon fluid are usually taken to be decoupled.
However, it is well known that in the early Universe many of these fluids
were coupled through annihilation and/or scattering processes \cite{wein}.
Many approachs to describe these primordial interactions have been proposed \cite{uzan,6,8,14,15,16}.
Furthermore, interactions between the two dark components have been investigated
in order to explain the coincidence problem, and effective phantom acceleration
without phantomic equation of state \cite{win,amen,varios,amen2}. Finally, interactions between two dark matter
components, or dark matter self-interaction \cite{ni}, and interaction between dark matter and baryons \cite{stein}
have also been investigated. In the last case, the possibility of interactions between baryons and dark matter 
due to the strong force
has been proposed because it can solve some conflicts between numerical simulations on structure formation 
with observations. One can justify the fact that such strong interactions
have not yet been detected because most of the WIMP detectors are ground based  \cite{stein}.

In this report, we explore the consequences of some phenomenological models of
interaction between two positive energy fluids with constant equation of state,
both satisfying the strong energy condition. We follow the line of research of 
Refs~\cite{varios,amen2}, postponing the physical justification of these phenomenological
proposals for future publications in order to concentrate on their physical consequences, 
and their potentialities to solve cosmological
puzzles (e.g, concordance problem, cosmological singularities),
with the hope that they may be tested soon \cite{dalai}. Assuming a power law form in the
scale factor for these interactions, similar to what was done in Ref.~\cite{amen2}
but not restricting to dark energy-dark matter interactions, we arrive at an effective Friedmann
equation containing two effective non interacting perfect fluids where one of them,
depending on the interaction, may have arbitrary equation of state, including
phantom-like behaviour, and/or negative energy density\footnote{Some similar situations
were studied in Refs.~\cite{cham}.}. This gives rise
to many types of cosmological analytical solutions which we obtain explicitely:
bouncing universes, as those studied in Refs.~\cite{ppn,fin,boz}, and
singular models with late
acceleration phases, like the $\Lambda$CDM model or others with phantomic
acceleration leading to a big-rip. In all these cases, the occurrence of interactions between
the fluids may lead, even in the framework of classical General Relativity, to accelerating
phases in the Universe, without the need to suppose that such fluids violate the strong
energy condition.

The paper is organized as follows: in Section II we describe our phenomenological
models and we arrive at the effective Friedmann equations, from which we obtain
the analytical solutions presented and discussed in Section III. Section IV is devoted
to the conclusions.

\section{Cosmology with two interacting fluids}

We model the interaction between two fluids satifying the strong energy condition
and with equations of state
$p_i=w_i \rho_i$, $i\in\{1,2\}$, with $w_i =$const.$>-1/3$, and $\rho_i>0$ as follows:
\begin{\en}
\dot{\r_{1}}+3H(1+w_1)\r_{1}&=&-Q , \nonumber \\
\dot{\r_{2}}+3H(1+w_2)\r_{2}&=&Q , \label{interacao}
\end{\en}
where $H=\dot{a}/a$ is the Hubble parameter, $Q$ is the interaction rate, and a
dot represents derivative in cosmic time.

We assume, without loss of generality, that the energy density of the second fluid
can be written as:
\begin{\eq}
\r_2=\r_{2_0}\frac{f}{f_0}\left( \frac{a_0}{a} \right)^{3(1+w_2)},
\label{materia}
\end{\eq}
where $\r_{2_0}$, $f_0$ and $a_0$ are constants, and $f$ is an
arbitrary time-dependent function. It follows from (\ref{interacao})
and (\ref{materia}) that
\begin{\eq}\label{Q}
Q=\r_2\frac{\dot{f}}{f}=\r_{2_0}\frac{\dot{f}}{f_0}\left( \frac{a_0}{a} \right)^{3(1+w_2)}
\end{\eq}
\par We now assign an ansatz for $f$,
\begin{\eq}
\label{f}
\frac{f}{f_0}=\left(\frac{a}{a_0} \right)^{-3w_f} ,
\end{\eq}
where $w_f$ is a constant. Hence
\begin{\eq}\label{materia2}
\r_2=\r_{2_0}\left(\frac{a}{a_0} \right)^{[-3(1+w_2+w_f)]}.
\end{\eq}
Consequently
\begin{\eq}\label{Q2}
Q=-3w_f \r_{2_0} H \left(\frac{a}{a_0} \right)^{-3(1+w_2+w_f)},
\end{\eq}
The solution of the conservation equation for the first fluid
\begin{\eq}
\dot{\r_{1}}+3H(1+w_1)\r_{1}=3w_f \r_{2_0}H \left(\frac{a}{a_0} \right)^{-3(1+w_2+w_f)}
\end{\eq}
is given by
\begin{\eq}
\label{r1}
\r_{1}=C_{1}a^{-3(1+w_1)}+\frac{\r_{2_0}w_f}{w_1-w_2-w_f}\left(\frac{a}{a_0} \right)^{-3(1+w_2+w_f)},
\end{\eq}
with $C_{1}$ a constant.
The total energy density then reads
\begin{\eq}
\label{rT}
\r_{T}=C_{1}a^{-3(1+w_1)}+\frac{\r_{2_0}(w_1-w_2)}{w_1-w_2-w_f}\left(\frac{a}{a_0} \right)^{-3(1+w_2+w_f)}.
\end{\eq}

Making the definitions,
\begin{\eq}
\label{def}
w_+\equiv w_1,\;\; w_- \equiv w_2+w_f, \;\; \r_+ \equiv C_{1}, \;\; \r_- = 
\frac{\r_{2_0}a_0^{3(1+w_2+w_f)}(w_+ - w_2)}{w_+ - w_-}
\end{\eq}
the Friedmann equation can be written as:
\begin{\eq}
H^2=l_{pl}^2\left(\r_{+}a^{-3(1+w_+)}+\r_- a^{-3(1+w_-)}\right),
\label{fried}
\end{\eq}
where $l_{pl}^2=8\pi G/3$.

Hence, from two interacting fluids satisfying the strong energy condition,
$w_1,w_2 > -1/3$, and positive energy density, $\r_1,\r_2 > 0$,
interacting via Eq.~(\ref{interacao}) with $Q$ given in (\ref{Q2}), one obtains two
effective non interacting fluids characterized by the parameters $w_+,w_-,\r_+,\r_-$
in the Friedmann equation (\ref{fried}), where one satisfies the strong energy
condition, $w_1 = w_+ > -1/3$, but the other can violate the strong energy condition
and/or have negative energy as long as,
from definitions (\ref{def}), and depending on the values of $w_f$ in
(\ref{Q2}), one may obtain $w_- < -1/3$ and/or $\r_- < 0$. This fact leads
to interesting cosmological solutions where bounces and/or late accelerated
phases may happen. We present these possibilities in the following section.
Note that if the two fluids have the same equation of state, $w_1=w_2$ there is no second effective fluid.
Also, if $w_+=w_-$, the quantity $\r_-$ cannot be defined. Hence we do not consider this singular
point in parameter space below.

\section{Analytical solutions}

By introducing a new coordinate time $\tau$:
\begin{\eq}
d \tau = \frac{d t}{a^\beta} \,, \quad
{\rm with} \quad \beta = \frac{3}{2} (2 w_+ - w_-+1)
\label{newtime}
\end{\eq}
we have as solution for the scale factor:
\begin{\eq}
a (\tau) = a_b \left(\frac{\tau^2}{\tau_0^2} - \frac{|\r_-|}{\r_-} \right)^\alpha \,,
\label{sfevolution}
\end{\eq}
with
\begin{eqnarray}
\alpha &=&  \frac{1}{3 (w_- - w_+)} \\
a_b &=& \left( \frac{|\rho_-|}{\rho_+} \right)^\alpha \\
\tau_0^2 &=& \frac{4\alpha^2}{l_{\rm pl}^2} \frac{|\rho_-|}{\rho^2_+}.
\end{eqnarray}

With the new coordinate time $\tau$, it is therefore possible
to generalize the solution in the conformal time found in \cite{fin},
and the cosmological models with
dust plus dark energy with arbitrary constant equation of state
obtained in \cite{GF}, to arbitrary values of $w_+ \,, w_-$.

The Hubble function is
\begin{\eq}
\label{hubble}
\frac{\dot a}{a}=\frac{2\alpha \tau}{\tau_0^2 a^{\beta}\left(\frac{\tau^2}{\tau_0^2}- \frac{|\r_-|}{\r_-}\right)},
\end{\eq}
while cosmic acceleration reads
\begin{\eq}
\label{acceleration}
\frac{\ddot a}{a}=-\frac{2\alpha}{\tau_0^2 a^{2\beta}\left(\frac{\tau^2}{\tau_0^2}- \frac{|\r_-|}{\r_-}\right)^2}
\left[(1+3w_+)\alpha\frac{\tau^2}{\tau_0^2} + \frac{|\r_-|}{\r_-}\right].
\end{\eq}
One may have transitions from decelerated to accelerated phases, and vice-versa, when
\begin{\eq}
\label{tautrans}
\tau^2=-\frac{|\r_-|}{\r_-}\frac{\tau_0^2}{\alpha(1+3w_+)}.
\end{\eq}

The asymptotic behaviours occur for $\tau\rightarrow\pm\infty$
and, if $\r_- >0$, for $\tau\rightarrow\pm\tau_0$. In the first case
one has
\begin{\eq}
t\propto |\tau| ^{3\alpha(1+w_+)},\; a(t) \propto |t|^{2/[3(1+w_+)]},
\label{tauinfinity}
\end{\eq}
while for the second one obtains, 
\begin{\eq}
t\propto |\tau \pm \tau_0|^{3\alpha(1+w_-)/2},\; a(t) \propto |t|^{2/[3(1+w_-)]},
\label{tautau0}
\end{\eq} 

Finally, looking at the evolution of the interaction term (\ref{Q2}), one should
expect that its modulus $|Q|$ would decrease when the Universe expands 
and Hubble time $t_H=1/H$ increases. which is indeed the case of Eq.~(\ref{Q2})
unless $w_-<-1\Rightarrow w_f<-1-w_2$. However, this extreme situation
can be explained using Eqs.~(\ref{materia2},\ref{r1}), where it is shown that the energy
densities of the two fluids $\rho_1$ and $\rho_2$ increase with expansion in that case, making the interaction stronger.
We will analyze these situations below in more details using the equation
\begin{eqnarray}
\label{qdot}
&-&\frac{a_b^{6(1+w_+)}}{3w_f \r_{2_0}a_0^{3(1+w_2+w_f)}}{\dot Q}=\nonumber\\
&-&\frac{2\alpha}{\tau_0^2 \left(\frac{\tau^2}{\tau_0^2}- \frac{|\r_-|}{\r_-}\right)^{2(1+w_-)/(w_--w_+)}}
\left[3\alpha(3+2w_- + w_+)\frac{\tau^2}{\tau_0^2} + \frac{|\r_-|}{\r_-}\right].
\end{eqnarray}

We will now discuss in more details the possible cases in terms of the
sign of $\r_-$. As we are assuming that both original
fluids satisfy the strong energy condition, then $w_+=w_1>-1/3$.

\subsection{$\r_- < 0$}

The scale factor reads:
\begin{\eq}
a(\tau) = a_b \left(\frac{\tau^2}{\tau_0^2}+1\right)^\alpha \, ,
\label{casoa}
\end{\eq}
implying that $-\infty<\tau<\infty$, and hence, from Eq.~(\ref{hubble}), the scale factor has
an extremum at $\tau=0$.

\subsubsection{$\alpha > 0$}

When $\alpha >0$, Eq.~(\ref{acceleration}) shows that this extremum is a minimum,
and we have a non singular bouncing universe. There are transitions from deceleration to acceleration
and vice-versa because there are real roots from Eq.~(\ref{tautrans}) in that case: $\alpha(1+3w_+)>0$.
As $w_->w_+$, the negative energy fluid dominates when the universe is small, avoiding the singularity
in the accelerated phase, while the positive energy fluid dominates
when the universe is large, which expands decelerately in this regime, as it can be seen 
from Eq.~(\ref{tauinfinity}).
 
Regarding the interaction term $Q$, one can see from Eq.~(\ref{qdot}) that for
\begin{\eq}
\label{tautransq}
0<\frac{\tau^2}{\tau_0^2}<\frac{1}{3\alpha(3+2w_-+w_+)}\approx 1
\end{\eq}
$Q$ is increasing while the model expands. This can be
understood by noticing that this is the period near the bounce\footnote{As $3+2w_-+w_+>2$ in this case,
$[3\alpha(3+2w_-+w_+)]^{-1}$ in Eq.~(\ref{qdot}) can be big only if $0<\alpha<<1$, but then the bounce also
lasts very long in $\tau$ (see Eq.~(\ref{tautrans}))}, where the scale
factor increases slowly while the Hubble time $t_h=1/H$ decreases abruptly, largely
compensating the slow expansion.

\subsubsection{$\alpha < 0$}

When $\alpha <0$, Eq.~(\ref{acceleration}) shows that the extremum is a maximum.
When the universe is small, the $w_+$ fluid dominates and
the $w_-$ fluid becomes important only around the maximum.
We have a big-bang big-crunch model (remember we are
assuming $w_+>-1/3$, hence $\tau\rightarrow\pm\infty$ implies $t$ finite in this case,
see Eq.~(\ref{tauinfinity})). There are no transitions from deceleration to acceleration
and vice-versa because $\alpha(1+3w_+)<0$ in this case (see Eq.~(\ref{tautrans})), whatever is
the value of $w_-$. 

The interaction term may increase with expansion only for the period
shown in Eq.~(\ref{tautransq}), now around the maximum of the scale factor, if $w_-<-(3+w_+)/2<-4/3$ 
(remember that $w_-<w_+$
in this case). This can be justified in terms of the slow increasing of $a$ in that period,
the increasing of the energy densities of the two fluids $\rho_1$ and $\rho_2$ because $w_-<-1$.

\subsection{$\r_- > 0$}

In this case the scale factor reads
\begin{\eq}
a(\tau) = a_b \left(\frac{\tau^2}{\tau_0^2}-1\right)^\alpha \, ,
\label{casob}
\end{\eq}
implying that $-\infty<\tau<-\tau_0$, or $\tau_0<\tau<\infty$.
Hence, from Eq.~(\ref{hubble}), the scale factor has no extremum being
either an always contracting or always expanding model. We will concentrate
on the always expanding solutions.

The conditions for having transitions from deceleration to acceleration
and vice-versa in this case are (see Eq.~(\ref{tautrans})), besides $\alpha(1+3w_+)<0$,
the condition that $|\tau|>|\tau_0|$ (see Eq.~(\ref{sfevolution})),
which implies that $|\alpha(1+3w_+)|<1$. Hence, in this case, $\alpha>0$
imposes that $w_+<-1/3$ and $w_->-1/3$, and $\alpha<0$
implies that $w_+>-1/3$ and $w_-<-1/3$, as it should be if the two effective fluids have positive
energy. As we are assuming $w_+>-1/3$, there are no transitions when $\alpha>0$.

\subsubsection{$\alpha > 0$}

When $\alpha >0$, both effective fluids satisfy the energy conditions and have positive energy.
These are the standard cases of two ordinary fluids governing a universe with decelerated expansion
from an initial singularity. The interaction term $Q$ always decreases with expansion.

\subsubsection{$\alpha < 0$}

When $\alpha <0$, the $w_+$ fluid dominates when the universe is small, necessarily reaching
a singularity (from Eq.~(\ref{tauinfinity}), $\tau\rightarrow\pm\infty$ implies $t$
finite in this case), and the $w_-$ fluid dominates when it is large. If $w_-<-1/3$ one can
have a transition from decelerated to accelerated expansion, and if $w_-<-1$, one gets a
big rip. 

As pointed above, in this last case one may have increasing $|Q|$ with expansion.
Looking at Eq.~(\ref{qdot}), one can see that, for $w_-\leq-(3+w_+)/2$, or $w_f\leq-(3+w_1+2w_2)/2$,
one has the bizarre situation where $|Q|$ always increases with expansion, even near the
initial singularity, when the ordinary $w_+=w_1$ fluid dominates and expansion takes place 
in the standard way. It seems that the increasing in $\r_2$ is so strong 
(see Eq.~(\ref{materia2})), that it compensates
expansion, and the decreasing of $\r_-$ and $H$.
When $-(3+w_+)/2<w_-<-1$ one has the more reasonable situation where
$|Q|$ decreases with expansion when the ordinary $w_+=w_1$ fluid dominates and increases only in
the period given by 
\begin{\eq}
\label{tautransq2}
1<\frac{\tau^2}{\tau^2_0}<-\frac{1}{3\alpha(3+2w_-+w_+)},
\end{\eq}
when we expect that the energy
densities of the two fluids $\rho_1$ and $\rho_2$ increase with expansion
(see Eqs.~(\ref{materia2},\ref{r1})), making the interaction stronger,
and Hubble time $t_H=1/H$ decreases.
Note that, depending on the value of $w_+$ and $w_-$, the increasing of $|Q|$ may happen
after $\r_- \r_+$ equilibrium, which takes place at $\tau^2 = 2 \tau_0^2$. For instance,
if $w_+=0$, there is a period when $|Q|$ decreases and the $w_-$ fluid dominates if
$-6/5<w_-<-1$. 

For $w_-\geq-1$, $|Q|$ always decreases with expansion.

\subsection{Final remarks}

For the sake of completeness, let us mention some different models one may obtain if the condition $w_+>-1/3$
is relaxed. One may have pre-big bang \cite{prebb} and singular inflationary models without graceful exit 
when $\rho_->0$. When $\rho_-<0$, there are
bouncing models between big rips or between accelerated contraction and expansion, and models which
expands from a pre-big bang like initial state, till a maximum size, and then contracts to a time 
reversed pre-big bang like final state, with transitions from accelerated to decelerated phases
and vice-versa around the maximum.
These are not realistic models.

From all these possibilities, the most interesting cases are the non singular solutions
and the models with late accelerated expansion. Among the first, there are
the bouncing models with $\r_- < 0$ and $\alpha>0$.
They are non singular models which are connected to a standard expansion phase
dominated by a fluid satisfying the strong energy condition when the universe is large.
In particular, the bouncing model studied
in Ref.~\cite{ppn} can be obtained with the choice $w_1=1/3$, $w_2=0$ and $w_f=1$, yielding
an interaction term given by $Q=-3KHa^{-6}$, which is strongly suppressed when $a$ is large.
One can view this interaction as happening in a temperature where baryons are relativistic but
dark matter is not, and they interact through the strong force as suggested in Ref.~\cite{stein}.
This could also happen in a temperature where both are relativistic, but not exactly with the same equation 
of state parameter. In that case on should have $w_f\approx 2/3$ and $Q\approx-2KHa^{-6}$.

In the second group there are the models with transition from decelerated to accelerated expansion,
as in the cases with parameters $\r_- > 0$, $\alpha<0$, and $w_-<-1/3$. One possibility would be to have
two different dark matter components, or dark matter and baryons, with $0<w_1<<1$, $0<w_2<<1$ 
and $w_1>w_2$, interacting with $w_f=-1$, yielding an effective
$\Lambda$CDM model with $w_+\approx 0$ and $w_-\approx -1$. In this case, one has $Q\approx3KH$, a mild
decreasing with expansion depending only on the Hubble parameter. Note that, as the effective cosmological
constant is given by $\Lambda\propto\rho_-\propto(w_1-w_2)$, its smallness could then be related with
the smallness of $w_1$ and $w_2$.

\section{Conclusion}

We have obtained some interesting cosmological models with flat spatial sections
from ordinary fluids which interact with the phenomenological interaction rate (\ref{Q2}).
Bouncing models,
which cannot be obtained from two non interacting fluids within classical
General Relativity unless one of them violates the null energy condition \cite{ppn2},
may arise in this context even if the two interacting fluids can be modelled by dust
and/or relativistic fluids, like dark matter components and baryons, depending on the
background temperature.
This same combination of fluids, with another choice of the power $w_f$ in Eq.~(\ref{Q2}),
may lead to the $\Lambda$CDM expansion rate without the need to introduce any fundamental
cosmological constant, whose small observational value is a challenge for particle physics theory.

Of course the relevance of the phenomenological interaction rate (\ref{Q2}) in different
phases of the history of the real Universe must be justified in micro-physical grounds. We will
come back to this problem in forthcoming publications. Also, the evolution of primordial
quantum perturbations in such models must be carried out in order to compare the results with
CMBR data. For instance, one should investigate if the bouncing models presented here could also lead to
scale invariant spectra of perturbations as it was verified in other bouncing universes \cite{ppp}.
These are subjects beyond the aim of the present report, which
is to point out that it is maybe not necessary to evoke the existence of exotic and/or unknown substances
to yield the present acceleration of the Universe and/or to avoid the initial singularity within classical 
General Relativity.


\begin{thebibliography}{99}

\bibitem{wein} S. Weinberg, {\it Gravitation and Cosmology}
(John Wiley and Sons, New York, 1972).

\bibitem{uzan} J. Uzan, Phys. Rev. {\bf 85}, 166 (1952).

\bibitem{6} J. Ehlers, "General Relativity and Kinetic Theory"
{\it General Relativity and Cosmology} Proc. of the Int. School of Physics
"Enrico Fermi", course XLVII (Academic Press, London/New York, 1971).

\bibitem{8} N. A. Chernikov, Acta Phys. Pol. {\bf 27},
465 (1964).

\bibitem{14} R. W. Lindquist, Ann. Phys. {\bf 37}, 487 (1966).

\bibitem{15} K. S. Thorne, Mont. Not. R. Astr. Soc. {\bf 194}, 439 (1981).

\bibitem{16} G. F. R. Ellis, D. R. Matravers and R. Treciokas, 
Ann. Phys. {\bf 150}, 455 (1983).

\bibitem{win} L. P. Chimento, A. S. Jakubi, D. Pavon and W. Zimdahl,
Phys. Rev. D {\bf 67} 083513 (2003); W. Zimdahl and D. Pavon,
Gen. Rel. Grav. {\bf 36}, 1483 (2004); W. Zimdahl, D. Pavon and L. P. Chimento,
Phys.Lett. B {\bf 521}, 133 (2001). 

\bibitem{amen} L. Amendola, Phys. Rev. D {\bf 62}, 043511 (2000);
L. Amendola and D. Tocchini-Valentini, Phys. Rev. D {\bf 64}, 043509, (2001);  
L. Amendola and C. Quercellini, Phys.Rev. D {\bf 68}, 023514 (2003). 

\bibitem{varios} H. Sadjadi and M. Allimohammadi, Phys.Rev. D {\bf 74}, 103007 (2006); 
R-G. Cai and A. Wang, JCAP {\bf 0503}, 002 (2005); 
A. de la Macorra, "The Fate of the Universe: Dark Energy Dilution?", arXiv:astro-ph/0701635;
Marek Szydlowski, Phys.Lett. B {\bf 632}, 1 (2006); 
M. Szydlowski, T. Stachowiak and R. Wojtak, Phys. Rev. D {\bf 73}, 063516 (2006).

\bibitem{amen2} E. Majerotto, D. Sapone, and L. Amendola, "Supernovae type Ia data favour coupled phantom energy", 
ArXiv astro-ph/0410543.

\bibitem{ni} M. Markevitch, A. H. Gonzalez, D. Clowe, A. Vikhlinin, L. David, W. Forman, C. Jones, S. Murray, and
W. Tucker, Astrophys. J. {\bf 606}, 819-824 (2004). 

\bibitem{stein} Adrienne L. Erickcek, Paul J. Steinhardt, Dan McCammon and Patrick C. McGuire, 
Phys.Rev. D {\bf 76}, 042007 (2007). 

\bibitem{dalai} N. Dalal, K. Abazajian, E. Jenkins and A. V. Manohar, 
Phys. Rev. Lett. {\bf 87}, 141302 (2001).

\bibitem{cham} L. P. Chimento, Phys. Lett. B {\bf 633}, 9 (2006). 

\bibitem{ppn} P.~Peter and N.~Pinto-Neto, Phys. Rev. D {\bf 66}, 063509 (2002).

\bibitem{fin} F.~Finelli, JCAP {\bf 0310}, 011(2003).

\bibitem{boz} V.~Bozza and G.~Veneziano, Phys.\ Lett.\ B {\bf 625}, 177 (2005).

\bibitem{prebb} G.~Veneziano, Phys. Lett. B {\bf 265}, 287 (1991);
M.~Gasperini and G.~Veneziano, Astropart. Phys. {\bf 1}, 317 (1993);
See also J.~E.~Lidsey, D.~Wands, and E.~J.~Copeland, Phys. Rep. {\bf
337}, 343 (2000) and G.~Veneziano, in {\sl The primordial Universe},
Les Houches, session LXXI, edited by P.~Bin\'etruy {\it et al.}, (EDP
Science \& Springer, Paris, 2000).

\bibitem{GF} A.~Gruppuso and F.~Finelli, Phys. Rev. D {\bf 73}, 023512 (2006).

\bibitem{ppn2} P.~Peter and N.~Pinto-Neto, Phys. Rev. D {\bf 65}, 023513 (2002).

\bibitem{ppp} P.~Peter, E.~Pinho and N.~Pinto-Neto, Phys. Rev. D {\bf 75},  023516 (2007). 

\end{thebibliography}
\end{document}